%% file: main.tex
\documentclass[11pt,a4paper,twoside]{article}
\usepackage[T1]{fontenc}
\usepackage[utf8]{inputenc}
\usepackage{amssymb}
\input{preamble}

\begin{document}

\title{Physics-Informed Neural Networks for Weakly Compressible Flows Using Galerkin-Boltzmann Formulation}
\author[1]{A.~Aygun\footnote{Corresponding author. E-mail address: atakana@metu.edu.tr} }
\author[1]{A.~Karakus}

\affil[1]{\textit{\small{Department of Mechanical Engineering, Middle East Technical University, Ankara, Turkey 06800}}}

\renewcommand\Authands{ and }
\date{\vspace{-5ex}}
\maketitle

\begin{abstract}
In this work, we study the Galerkin-Boltzmann formulation within a physics-informed neural network (PINN) framework to solve flow problems in weakly compressible regimes. 
The Galerkin-Boltzmann equations are discretized with second-order Hermite polynomials in microscopic velocity space, which leads to a first-order conservation law with six equations. Reducing the output dimension makes this equation system particularly well suited for PINNs compared with the widely used D2Q9 lattice Boltzmann velocity space discretizations.  We created two distinct neural networks to overcome the scale disparity between the equilibrium and non-equilibrium states in collision terms of the equations. We test the accuracy and performance of the formulation with benchmark problems and solutions for forward and inverse problems with limited data. Our findings show the potential of utilizing the Galerkin-Boltzmann formulation in PINN for weakly compressible flow problems.

\textbf{Keywords:} Galerkin formulation, Boltzmann equation, weakly compressible flow, physics-informed neural networks 
\end{abstract}

\section{Introduction}
\label{sec:introduction}
\input{introduction.tex}

\section{Methodology}
\label{sec:methodology}
\input{methodology}

\section{Numerical Tests}
\label{sec:results}
\input{results.tex}

\section{Conclusion}
\label{sec:conclusion}
\input{conclusion}

\bibliographystyle{ieeetr}
\bibliography{main}

\end{document}

%% file: preamble.tex
\usepackage{lipsum}
\usepackage{authblk}
\usepackage[top=3cm, bottom=3cm, left=2cm, right=2cm]{geometry}
\usepackage{fancyhdr}
\usepackage{adjustbox}
\usepackage[sort&compress, numbers, compress ]{natbib}
\usepackage[colorlinks,bookmarksopen,bookmarksnumbered,citecolor=red,urlcolor=red]{hyperref}
\usepackage{algorithm}
\usepackage{algorithmic}
\usepackage{boxedminipage}
\usepackage{comment}
\usepackage{eqparbox}
\usepackage{multirow}
\usepackage{multicol}
\usepackage{caption}
\usepackage{subcaption}

\usepackage{nomencl}
\makenomenclature

\usepackage{amsmath}
\usepackage{amsfonts}

\usepackage{graphicx}

\usepackage{tikz}

\graphicspath{{./figures}}

\newlength{\commentindent}
\usepackage{array}
\newcolumntype{R}[1]{>{\raggedleft\let\newline\\\arraybackslash\hspace{0pt}}m{#1}}

\newcommand\BibTeX{{\rmfamily B\kern-.05em \textsc{i\kern-.025em b}\kern-.08em
T\kern-.1667em\lower.7ex\hbox{E}\kern-.125emX}}

\newcommand{\grad}{\nabla}

\newcommand{\bx}{\mathbf{x}}
\newcommand{\bu}{\mathbf{u}}

\newcommand{\feq}{f^{eq}}

\newcommand{\qeq}[1]{q_{#1}^{eq}}
\newcommand{\qneq}[1]{q_{#1}^{neq}}

\let\OLDthebibliography\thebibliography
\renewcommand\thebibliography[1]{
	\OLDthebibliography{#1}
	\setlength{\parskip}{0pt}
	\setlength{\itemsep}{0pt plus 0.3ex}
}

\renewenvironment{abstract}{%
	\begin{center}
		\begin{adjustbox}{minipage=0.90\textwidth,center} 
			\rule{\textwidth}{0.5pt}}
		{\par\noindent\rule{\textwidth}{0.5pt}  
		\end{adjustbox}
	\end{center} 
}

\makeatletter

\renewcommand\@maketitle{%
	\begin{adjustbox}{minipage=0.90\textwidth,center}
		\begin{center}
			\vskip 1em
			\let\footnote\thanks 
			{\LARGE \@title \par }
			\vskip 1.0em
			{\large \@author \par}
		\end{center}
	\end{adjustbox}
	\vskip 0.0em \par
}
\makeatother
\usepackage{etoolbox}
\renewcommand\nomgroup[1]{%
  \item[\bfseries
  \ifstrequal{#1}{G}{Greek Symbols}{%
  \ifstrequal{#1}{S}{Symbols}{%
  \ifstrequal{#1}{A}{Abbreviations}{%
  \ifstrequal{#1}{B}{Subscripts}{%
  \ifstrequal{#1}{P}{Superscripts}{
  }}}}}]}

%% file: introduction.tex
The Boltzmann equations describe the time evaluation of the particle distribution based on the kinetic theory \citep{chattopadhyay_fluid_2023}. In the low Mach limit, these equations recover the Navier-Stokes equations \citep{cercignani_boltzmann_1988}. Boltzmann equations with the Bhatnagar-Gross-Krook (BGK) collision model \citep{bhatnagar_model_1954} have been solved using numerical methods. Lattice Boltzmann method is widely used \citep{yu_viscous_2003, lallemand_lattice_2021, wissocq_hydrodynamic_2022} due to its simple formulation and application in fluid flow problems as well as suitability for parallel computing \citep{aidun_lattice-boltzmann_2010}. Moreover, the finite volume and discontinuous Galerkin methods are also used to solve the Boltzmann equations \citep{min_spectral-element_2011, li_high_2017, karakus_discontinuous_2019}.

Recent advancements in deep learning algorithms have demonstrated their potential in solving partial differential equations (PDE) \cite{lee1990neuralAlgo, lagaris1998artificial, sirignano_dgm_2018}. Raissi et al. \citep{raissi_physics-informed_2019} introduced the concept of physics-informed neural networks (PINN) to approximate the solution of PDEs by tuning the parameters of the neural network. This approach uses the underlying physical law from the differential equation alongside the known values on the boundary conditions and initial conditions. By utilizing automatic differentiation \citep{baydin_automatic_2017}, PINNs can approximate the required derivatives without the need for traditional meshing, contrary to conventional numerical methods.

Despite the success of PINNs across various problems, recent studies have highlighted the challenges in training these networks to obtain accurate solutions. PINNs can struggle with multiscale and multiphysics problems \cite{karniadakis_physics-informed_2021}, and fully connected networks often face difficulties in learning high-frequency functions. Wang et al. \citep{wang_when_2022} analyze this phenomenon named spectral bias in the lens of neural tangent kernel \cite{jacot_neural_2018}. Wang et al. examine the disparity in convergence rates among the various terms defining a PINN loss function \cite{wang_understanding_2021}. To address this issue, Wang et al. \cite{wang_eigenvector_2021} introduce Fourier feature embedding for input coordinates, which helps in capturing multiscale behavior. Several other studies have focused on balancing the influence of individual loss terms to prevent any single term from dominating. In \cite{mcclenny_self-adaptive_2023}, the authors allow the weights to be a trainable parameter in the learning process to obtain self-adaptive PINNs. Anagnostopoulos et al. \cite{anagnostopoulos_residual-based_2024} develop a gradient-less weighting scheme by deriving weights from the residual values of individual points within the framework of evolving cumulative residuals. Furthermore, in \cite{wang_respecting_2024}, the training procedure was modified to respect the causality of physical systems, particularly in time-dependent problems. In this approach, the weights are calculated from the cumulative residual loss of previous time steps.

There have been some studies solving the Boltzmann equation using PINN. Lou et al. \citep{lou_physics-informed_2021} try to solve forward and inverse problems with the discrete velocity Boltzmann model using the BGK collision model. They employ D2Q9 and D2Q16 formulations in two-dimensional problems. In \cite{li_physics-informed_2021, zhou_physics-informed_2023}, the authors solve the mode-resolved phonon Boltzmann transport equation for thermal transport problems. Oh et al. \citep{oh_separable_2024} use separable PINN with the integration of Gaussian functions and a relative loss approach. These improvements enhance the accuracy of macroscopic moment approximations. 

In this study, we integrate the Galerkin-Boltzmann formulation \citep{tolke_discretization_2000, karakus_discontinuous_2019} within a physics-informed neural network framework to develop a solution strategy for flows in weakly compressible regimes. The Galerkin-Boltzmann equations are discretized with second-order Hermite polynomials in microscopic velocity space, and the collision term is modeled with the BGK model. This methodology yields a first-order conservation law with six equations. Compared with the widely used D2Q9 lattice Boltzmann velocity space discretization, this formulation reduces the output dimension. These properties make this formulation well-suited for PINN architectures. However, there is a scale difference between the advection term and the collision operator, leading to converge issues in PINN training. Therefore, in the PINN architecture, two neural networks are present to overcome this problem. The first neural network predicts the primitive flow variables used to calculate the equilibrium state, while the other network predicts the variables in the non-equilibrium state. Additionally, we used Fourier feature embeddings in the input layer to accelerate convergence. 

The remainder of this paper is organized as follows. In Section \ref{sec:methodology}, we introduce the Galerkin-Boltzmann formulation and give some preliminaries on PINN. We also provide details about the network architecture tailored to solve the Boltzmann system. In section \ref{sec:results}, we present the results obtained by Galerkin-Boltzman formulation using PINN for steady and transient problems with exact solutions followed by a solution obtained with limited data. Finally, in Section \ref{sec:conclusion}, we make some concluding remarks.

%% file: methodology.tex
In this section, we begin by introducing the Galerkin-Boltzmann formulation. We then explain the modified physics-informed neural network structure to solve these Boltzmann equations.

\subsection{Galerkin-Boltzmann Formulation}
We consider the Boltzmann equations with the BGK collision model \cite{bhatnagar_model_1954}. This formulation describes the time evolution of a particle distribution function $f(\bx, \mathbf{v}, t)$, which is a function of spatial variable $\bx$, microscopic velocity $\mathbf{v}$ and time $t$. The equation can be written as

\begin{equation}
    \label{eq:boltzmann_bgk}
    \frac{\partial f}{\partial t} + \mathbf{v} \cdot \grad f = \frac{\feq - f}{\tau}
\end{equation}
where $\tau$ is the relaxation time, which is a function of dynamic viscosity, and $\feq$ is the equilibrium state distribution function, which can be written as 

\begin{equation}
    \label{eq:f_equilibrium}
    \feq = \frac{\rho}{2\pi RT} \exp\left(-\frac{(\mathbf{v}-\bu)^2}{2RT}\right)
\end{equation}
where $\rho, \:R, \: T, \: \mathbf{v}$ are macroscopic density, gas constant, temperature, and macroscopic velocity, respectively.

To obtain the Galerkin-Boltzmann formulation, we follow the formulation of Karakus et al. \cite{karakus_discontinuous_2019}, which originated from the formulation of T\"olke et al. \cite{tolke_discretization_2000}. We approximate the phase space distribution function by a polynomial expansion,

\begin{equation}
    \label{eq:polynomial_expansion}
    f(\bx, \mathbf{v}, t) = \frac{1}{2\pi RT} \exp \left(-\frac{|\mathbf{v}|^2}{2RT} \right) \sum_{n=1}^{N_h} q_n(\bx, t) \phi_n(\mathbf{v})
\end{equation}
where $\phi_n(\mathbf{v})$ are the polynomials in the velocity space and $q_n$ are their corresponding coefficients. We approximate the equilibrium phase space density $\feq$ with the unknown coefficients $\qeq{n}$. A system for the unknown coefficients can be obtained by substituting these polynomial approximations into Equation (\ref{eq:boltzmann_bgk}). We follow the Galerkin formulation and multiply the resulting equation with some test functions from the same polynomial space, $\{ \phi_m(\mathbf{v})\}^{N_h}_{m=1}$, and integrate over the unbounded microscopic velocity space $\Omega_{\mathbf{v}} = (-\infty, \infty)^d$ to obtain,

\begin{equation}
    \label{eq:galerkin_formulation}
    \int_{\Omega_{\mathbf{v}}} \phi_m(\mathbf{v}) \frac{1}{2\pi RT}\exp \left(-\frac{|\mathbf{v}|^2}{2RT} \right) \phi_n(\mathbf{v}) \left(  \frac{\partial q_n}{\partial t} + \mathbf{v} \cdot \grad q_n - \frac{\qeq{n} - q_n}{\tau}\right) \, d{\mathbf{v}} = 0
\end{equation}

By comparing this expression with \ref{eq:boltzmann_bgk}, we see that $\frac{1}{2\pi RT}\exp \left(-\frac{|\mathbf{v}|^2}{2RT} \right) \phi_n(\mathbf{v}) \qeq{n}$ is a weighted $L_2$ approximation of the local equilibrium distribution. In other words, the coefficients $\qeq{n}$ satisfy

\begin{equation}
    \label{eq:boltzmann_integral}
    \int_{\Omega_{\mathbf{v}}} \phi_m(\mathbf{v}) \left( \frac{1}{2\pi RT} \exp \left( - \frac{|\mathbf{v}|^2}{2RT} \right) \phi_n \qeq{n} - \feq \right) \, d\mathbf{v} = 0.
\end{equation}
We can then write the system in Equation (\ref{eq:galerkin_formulation}) in matrix notation as 

\begin{equation}
    \label{eq:matrix_form}
    M_{mn} \frac{\partial q_n}{\partial t} = A_{\bx,mn} \cdot \grad_{\bx}q_n + M_{mn} \frac{1}{\tau} \left( \qeq{n} - q_n\right),
\end{equation}
where the velocity space mass matrix, $M$, and stiffness matrices $A_{\bx}$ can be written as
\begin{equation}
    \label{eq:mass_matrix}
    M_{mn} = \frac{1}{2\pi RT} \int_{\Omega_{\mathbf{v}}} \phi(m)\phi(n)\exp \left(- \frac{|\mathbf{v}|^2}{2RT}\right) \, d\mathbf{v},
\end{equation}
\begin{equation}
    \label{eq:stiffness_matrix}
    A_{\bx,mn} = - \frac{1}{2\pi RT} \int_{\Omega_{\mathbf{v}}} \phi(m)\phi(n)\mathbf{v}\exp \left(- \frac{|\mathbf{v}|^2}{2RT}\right) \, d\mathbf{v}.
\end{equation}
The exponential weighting leads us to use Hermite polynomials to approximate the velocity space. Using the bi-variate Hermite polynomials, which are orthonormal under the Maxwellian inner product, we get some desirable properties: the mass matrix is the identity matrix, and stiffness matrices become constant. Using Equation (\ref{eq:boltzmann_integral}), the unknown equilibrium coefficients $\qeq{n}$ can be linked to $q_n$ explicitly. Then, we can obtain the first-order partial differential equation as 

\begin{equation}
    \label{eq:bns_pde}
    \frac{\partial q}{\partial t} = A_{\bx} \cdot \nabla_{\bx}q + \mathcal{N}(q),
\end{equation}
where $q = q(\bx, t)$ is a vector of Hermite polynomial coefficients, $A_{\bx}$ are directional coefficient matrices, and $\mathcal{N}$ is the collision operator. The order of the polynomial space must be sufficiently large to recover the macroscopic flow properties. Using second-order polynomial space to approximate the phase space in a spatial dimension of two yields a vector of unknown polynomial coefficients as $q(\bx,t) = [q_1(x, y, t), \dots\ q_6(x, y, t)]$ and the operators $A_{\bx} = [A_x, A_y]$, given by 

\begin{eqnarray*}
A_x = -\sqrt{RT} \left(\begin{array}{cccccc}
0 & 1 & 0 & 0 & 0 & 0\\
1 & 0 & 0 & 0 & \sqrt{2} & 0\\
0 & 0 & 0 & 1 & 0 & 0\\
0 & 0 & 1 & 0 & 0 & 0\\
0 & \sqrt{2} & 0 & 0 & 0 & 0\\
0 & 0 & 0 & 0 & 0 & 0
\end{array}
\right),\;
A_y = -\sqrt{RT} \left(\begin{array}{cccccc}
0 & 0 & 1 & 0 & 0 & 0\\
0 & 0 & 0 & 1 & 0 & 0\\
1 & 0 & 0 & 0 & 0 & \sqrt{2} \\
0 & 1 & 0 & 0 & 0 & 0\\
0 & 0 & 0 & 0 & 0 & 0\\
0 & 0 & \sqrt{2} & 0 & 0 & 0
\end{array}
\right) 
\end{eqnarray*}
and the collision operator is given by 

\begin{eqnarray}
\label{eq:nonlinearDef}
\mathcal{N} = -\frac{1}{\tau}\left(\begin{array}{cccccc}
0 &
0 &
0 &\
\left(q_4 - \frac{q_2q_3}{q_1}\right) &
\left(q_5 - \frac{q_2^2}{q_1\sqrt{2}}\right) &
\left(q_6 - \frac{q_3^2}{q_1\sqrt{2}}\right)
\end{array}
\right)^T,
\end{eqnarray}
where $c = \sqrt{RT}$ represents the speed of sound. This Galerkin-Boltzmann system of (\ref{eq:bns_pde}) recovers the Navier-Stokes equations for low Mach number, nearly incompressible flows \citep{karakus_discontinuous_2019}. Kinematic viscosity can be calculated as $\nu = \tau RT$. The relations between macroscopic flow properties and moment of the distribution functions can be written as
\begin{eqnarray*}
\rho = q_1, \; \rho u = \sqrt{RT}q_2, \;\rho v = \sqrt{RT}q_3.
\end{eqnarray*}
The non-equilibrium state contribution to the first three fields is zero. Therefore,
\begin{eqnarray*}
 \qeq{1} = q_1, \; \qeq{2} = q_2, \;\qeq{3} = q_3.
\end{eqnarray*}
The other three equilibrium state contributions can be written as the combination of the first three fields as
\begin{eqnarray*}
\qeq{4} = \frac{q_2 q_3}{q_1}, \; \qeq{5} = \frac{q_2^2}{q_1\sqrt{2}}, \;\qeq{6} = \frac{q_3^2}{q_1\sqrt{2}}.
\end{eqnarray*}
%
%
%
The pressure is recovered by the equation of state for ideal gases $p = \rho RT$.

\subsection{Physics-Informed Neural Networks}
For the physics-informed neural network, we consider the general form of partial differential equations:
\begin{subequations}
\label{eq:PDEsystem}
\begin{align}
 &\mathbf{u}_t + \mathcal{N}[\bu] = 0, \quad &\mathbf{x} \in \Omega,\; t\in[0,T]\label{eq:PDE}\\
 &\mathbf{u}(\mathbf{x},0) = f(\mathbf{x}), \quad &\mathbf{x} \in \Omega\label{eq:PDE IC}\\
 &\mathcal{B}[\bu] = 0, \quad &\mathbf{x} \in \partial\Omega, \; t \in [0,T]\label{eq:PDE BC}
\end{align}
\end{subequations}
where $\mathcal{N}$ is a generalized differential operator, $\bx \in \mathbb{R}^d$ and $t \in [0, T]$ are the spatial and temporal coordinates. $\Omega \;\text{and}\; \partial\Omega$ represent the computational domain and the boundary respectively. $\mathbf{u}(\bx,t)$ is the general solution of the PDE with $f(\bx)$ as the initial condition and $\mathcal{B}[\cdot]$ denotes the boundary operator that can be various boundary conditions such as Dirichlet, Neumann, Robin, and periodic boundary conditions. We approximate the solution $\bu(\bx, t)$ by the PINN formulation of Raissi et al. \citep{raissi_physics-informed_2019} by a feedforward neural network $\hat{\bu}(\bx, t; \theta)$ where $\theta$ denotes the hyperparameters of the network, weights and biases. All required gradients with respect to input variables are computed with automatic differentiation \citep{baydin_automatic_2017} and hence we can define the PDE residual as 
\begin{eqnarray}
    \label{eq:pde_loss}
    \mathcal{R}_{pde}(\bx, t) = \hat{\bu}_t + \mathcal{N}[\hat{\bu}] = 0, \quad &\mathbf{x} \in \Omega,\; t\in[0,T]
\end{eqnarray}
and the boundary and initial residuals can be defined, respectively, by
\begin{eqnarray}
    \label{eq:boundary_loss}
    \mathcal{R}_{bc}(\bx, t) = \mathcal{B}[\hat{\bu}](\bx, t), \quad &\mathbf{x} \in \partial\Omega, \; t \in [0,T]
\end{eqnarray}
and
\begin{eqnarray}
    \label{eq:initial_loss}
    \mathcal{R}_{ic}(\bx) = \hat{\bu}(\bx, 0) - f(\bx), \quad &\mathbf{x} \in \Omega.
\end{eqnarray}
Then, we train the network by minimizing a composite loss function,
\begin{eqnarray}
    \label{eq:composite_loss}
    \mathcal{L} = w_r\mathcal{L}_r + w_{bc}\mathcal{L}_{bc} + w_{ic}\mathcal{L}_{ic}.
\end{eqnarray}
Here, $L_r$ is the PDE residual loss, $L_{bc}$ is the boundary loss, and $L_{ic}$ is the initial condition loss. The $w$ terms are specifically assigned weights of each loss term, which can be tuned manually or adaptively \cite{wang_understanding_2021, wang_when_2022, mcclenny_self-adaptive_2023, anagnostopoulos_residual-based_2024}. Each loss term can be written as 
\begin{subequations}
\begin{align}
    &\mathcal{L}_r = \frac{1}{N_r} \sum_{i=1}^{N_r} \big| \mathcal{R}_{pde}(\bx_r^i, t_r^i) \big|^2 \\
    &\mathcal{L}_{bc} = \frac{1}{N_{bc}} \sum_{i=1}^{N_{bc}} \big| \mathcal{R}_{bc}(\bx_{bc}^i, t_{bc}^i) \big|^2 \\
    &\mathcal{L}_{ic} = \frac{1}{N_{ic}} \sum_{i=1}^{N_{ic}} \big| \mathcal{R}_{ic}(\bx_{ic}^i) \big|^2.
\end{align}
\end{subequations}
Here $N_r, \; N_{bc}$, and $N_{ic}$ are the total number of points used for calculating the mean squared error used here as the loss function.

\begin{figure}
    \centering
    \includegraphics[width=0.9\textwidth]{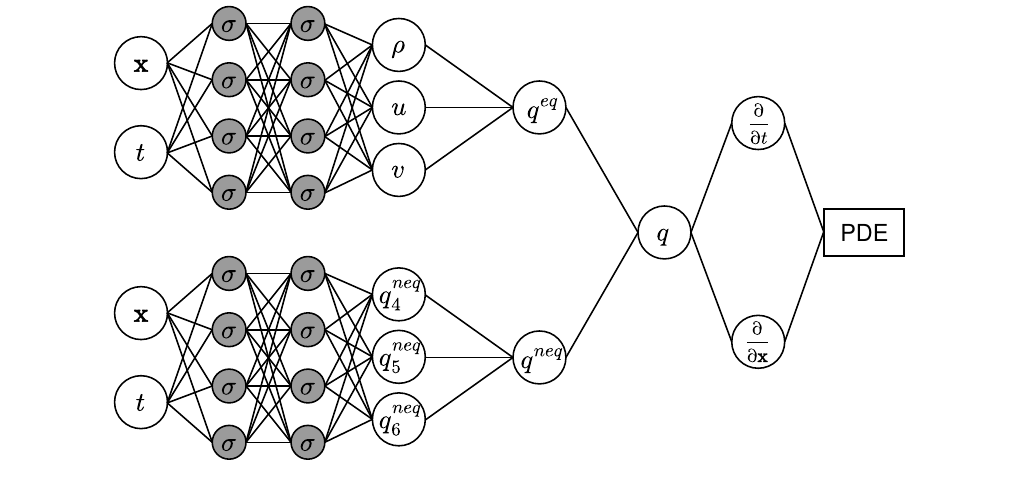}
    \caption{Network structure for solving Galerkin-Boltzmann formulation shown for a two-dimensional case.}
    \label{fig:pinn_figure}
\end{figure}

In this work, we train two separate networks for the equilibrium and non-equilibrium state variables similar to \cite{lou_physics-informed_2021, li_physics-informed_2021, zhou_physics-informed_2023}. The network structure can be seen in Figure \ref{fig:pinn_figure}. Both networks take spatio-temporal coordinates as the input. The equilibrium network outputs the primitive flow variables, velocity, and density. Using these primitive flow variables, the equilibrium state variables $\qeq{n}(\bx,t) = [\qeq{1}(\bx, t), \dots\ , \qeq{6}(\bx, t)]$ can be calculated. The second network outputs the non-equilibrium state variables. The number of non-equilibrium state variables is three, as it can be seen in Equation (\ref{eq:nonlinearDef}) since $\qneq{1} = \qneq{2} = \qneq{3} = 0$. We also scale the output of the second network with the relaxation time parameter $\frac{1}{\tau}$; otherwise, the values are close to zero, and it does not converge to physically admittable results. The boundary conditions are employed by using the primitive flow variables in the first neural network with equilibrium state variables. The non-equilibrium network accepts boundary conditions using Equation (\ref{eq:bns_pde}). 

We have also used Fourier feature embedding to enrich the input space \citep{tancik_fourier_2020, wang_eigenvector_2021}. The input coordinates are mapped into a higher-dimensional feature space using random Fourier mapping $\gamma$ as 

\begin{equation}
    \gamma (\bx) = \begin{bmatrix}
        \cos (\mathbf{B}\bx) \\ \sin (\mathbf{B}\bx)
    \end{bmatrix},
\end{equation}
where all the entries in $\mathbf{B}$ is sampled from a Gaussian distribution $\mathcal{N}(0, \sigma^2)$ with the standard deviation $\sigma$ being a user-specified parameter.

%% file: results.tex
In this section, the results of the Galerkin-Boltzmann formulation are presented. The PINN code is written with PyTorch as the backend. We select ADAM optimizer and hyperbolic tangent activation function for all the test cases presented. All the results shown here are obtained using a Dell Precision laptop with an Nvidia RTX A1000 laptop GPU. 

\subsection{Kovasznay Flow}
We first test our formulation to simulate a two-dimensional steady Kovasznay flow \cite{kovasznay1948laminar}. The rectangular domain is between $-0.5 \leq x \leq 2$ and $-0.5 \leq y \leq 1.5$. The exact solutions are given as 

\begin{subequations}
\label{eq:kovasznay_exact}
\begin{align}
 & u(x,y) = u_0 \left( 1 - \exp(\lambda x) \cos(2\pi y)\right),\\
 & v(x,y) = u_0 \left( \frac{\lambda}{2\pi} \exp(\lambda x) \sin(2\pi y) \right), 
\end{align}
\end{subequations}
where $u$ and $v$ represents the fluid velocity and 

\begin{equation}
    \lambda = \frac{Re}{2} - \sqrt{\frac{Re^2}{4} + 4\pi ^2}.
\end{equation}
We employ the test case from \cite{lou_physics-informed_2021}, where they have used a two-dimensional-nine-speed (D2Q9) model as a lattice-Boltzmann method. The flow variables are $Re=10$, $u_0=0.1581$, $p_0=0.05$, $RT=100$, $c=17.3205$ which makes $Ma=0.0091$. This enables us to stay in the weakly compressible regime. Using these values, we get the kinematic viscosity as $\nu = 0.0158$ and the relaxation parameter as $\tau = 1.58 \times 10^{-4}$. The boundary conditions are given using Equation (\ref{eq:kovasznay_exact}) for the velocities. The density boundary condition is given as $\rho=1$ in all boundaries. 

\begin{figure}
    \centering
    \begin{subfigure}[b]{\textwidth}
        \includegraphics[width=\textwidth]{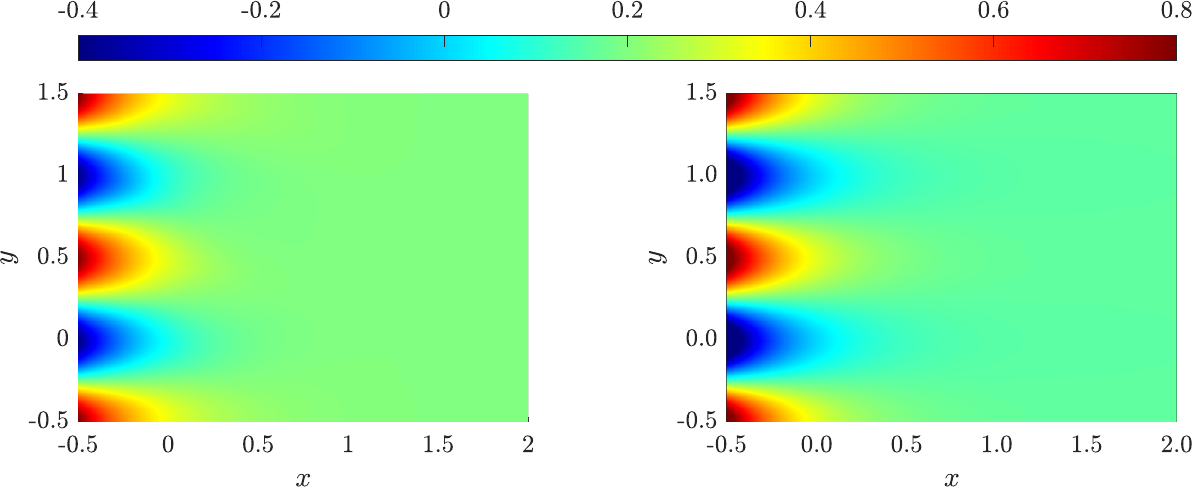}
        \caption{$u$ velocity contour.}
    \end{subfigure}
    ~
    ~
    \begin{subfigure}[b]{\textwidth}
        \includegraphics[width=\textwidth]{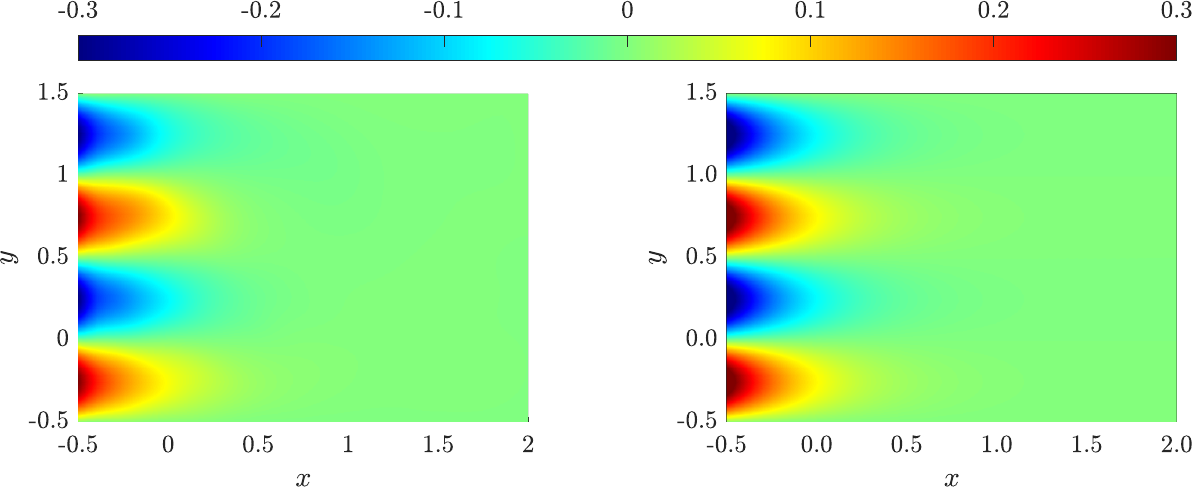}
        \caption{$v$ velocity contour.}
    \end{subfigure}
    ~
    \caption{Velocity contours for Galerkin-Boltzmann solutions for the Kovasznay flow. The top row shows the $u$ velocity contour, and the bottom row shows the $v$ velocity contour. The left column shows the prediction by PINN and the right column is the exact solution.}
    \label{fig:kovasznay_u_v}
\end{figure}

We have used 4 layers with 128 neurons in both equilibrium and non-equilibrium networks for the network architecture. Both networks are initialized with the Glorot scheme \citep{glorot_understanding_2010} and trained for 20000 epochs. The learning rate starts at 0.005 and exponentially decays with a decay rate of 0.99 for the ADAM optimizer. The PINN solution with the exact solutions for $u$ and $v$ velocities can be seen in Figure \ref{fig:kovasznay_u_v}. We can accurately capture the flow field using PINN with the Galerkin-Boltzmann formulation compared to the exact solution.

We also solved the same problem using incompressible Navier-Stokes (INS) equations with PINN. The flow parameters are the same as the solution in the Galerkin-Boltzmann formulation. We employed only one network with the same number of layers and neurons (4 hidden layers with 128 neurons) to get the primitive flow variables of pressure and velocities with the Fourier feature embedding. The number of epochs for the training is again 20000. 

\begin{figure}
    \centering
    ~
    \begin{subfigure}[b]{\textwidth}
        \includegraphics[width=\textwidth]{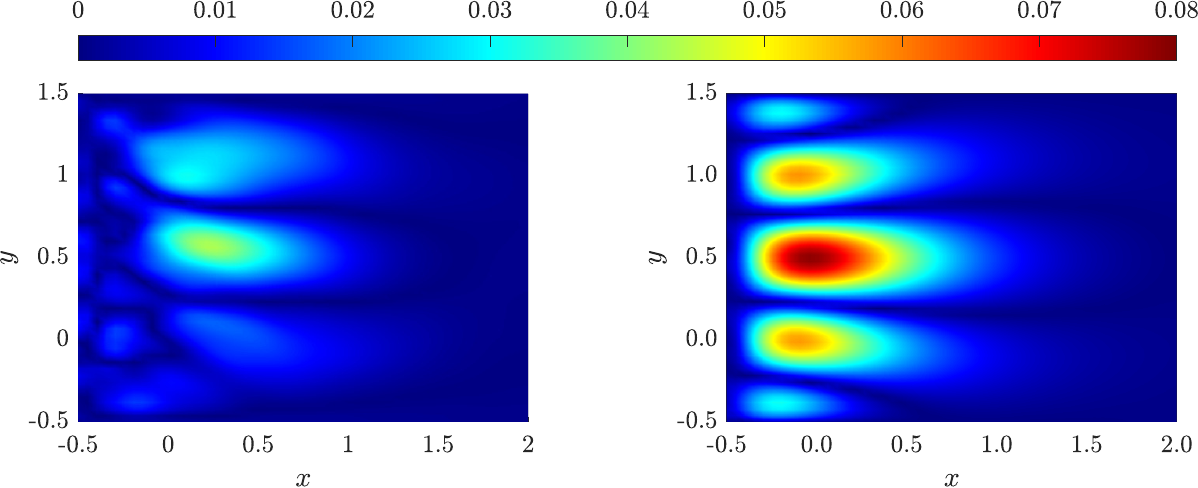}
        \caption{$u$ velocity error}
    \end{subfigure}
    ~
    ~
    \begin{subfigure}[b]{\textwidth}
        \includegraphics[width=\textwidth]{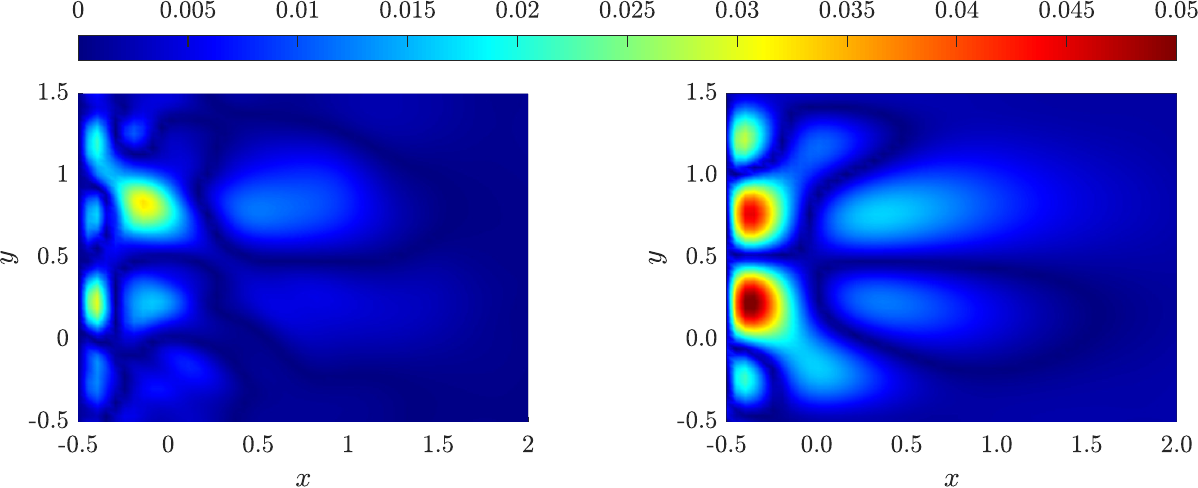}
        \caption{$v$ velocity error}
    \end{subfigure}
    ~
    \caption{Velocity error contours for Galerkin-Boltzmann and INS solutions of the Taylor-Green vortex. The top row shows the $u$ velocity error, and the bottom row shows the $v$ velocity error.}
    \label{fig:velocity_error}
\end{figure}

The error contours for the velocities for both Galerkin-Boltzmann and INS solutions are presented in Figure \ref{fig:velocity_error}. As can be seen from the figure, PINN can predict a more accurate solution with the Galerkin-Boltzmann formulation under these training settings than INS. Especially INS formulation struggles more to capture the true solution near the boundaries without any special treatment. We have also presented the $L_2$ errors for $u$ and $v$ velocity solutions under both formulations in Table \ref{tab:boltzmann_vs_ins} with the total training time. Although the Galerkin-Boltzmann formulation uses two neural networks, hence it has more hyperparameters, the training time is comparable with the INS formulation. This can result from the special structure of the PDE system, which uses only first-order derivatives.  

\begin{table}
    \centering
    \caption{$L_2$ error of $u$ and $v$ velocities and training time comparison of Galerkin-Boltzmann formulation and INS for the solution of Kovasznay flow with PINN.}
    \begin{tabular}{ccccccc}
    \hline \hline
        Formulation & & $Err_u$ &  &  $Err_v$ & & Training time (min) \\ \hline
        Galerkin-Boltzmann& & 0.049 & & 0.028 & & 32.47 \\ \hline 
         INS & & 0.096 & & 0.044 & & 30.30 \\ \hline \hline
    \end{tabular}
    \label{tab:boltzmann_vs_ins}
\end{table}

\subsection{Taylor-Green Vortex}
We solve the Taylor-Green vortex problem, which has a time-dependent exact solution to test the formulation in unsteady problems. The computational domain is $-\pi \leq x \leq \pi$ and $-\pi \leq y \leq \pi$. The exact solution is given as 

\begin{subequations}
    \label{eq:tgv_exact}
    \begin{align}
        & u(x,y,t) = -\cos(x) \sin(y) \exp(-2t\nu), \\
        & v(x,y,t) = \sin(x) \cos(y) \exp(-2t\nu),
    \end{align}
\end{subequations}
where $u$ and $v$ are the $x$ and $y$ components of fluid velocity respectively. 

\begin{figure}
    \centering
    \includegraphics[width=.45\linewidth]{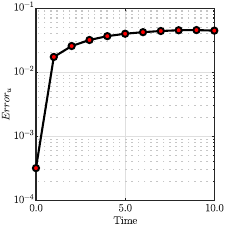}
    \caption{Evaluation of error norm of the u velocity through time for Taylor-Green flow.}
    \label{fig:Error_u_tgv}
\end{figure}

\begin{figure}
    \centering
    \begin{subfigure}[b]{0.45\textwidth}
        \includegraphics[width=\textwidth]{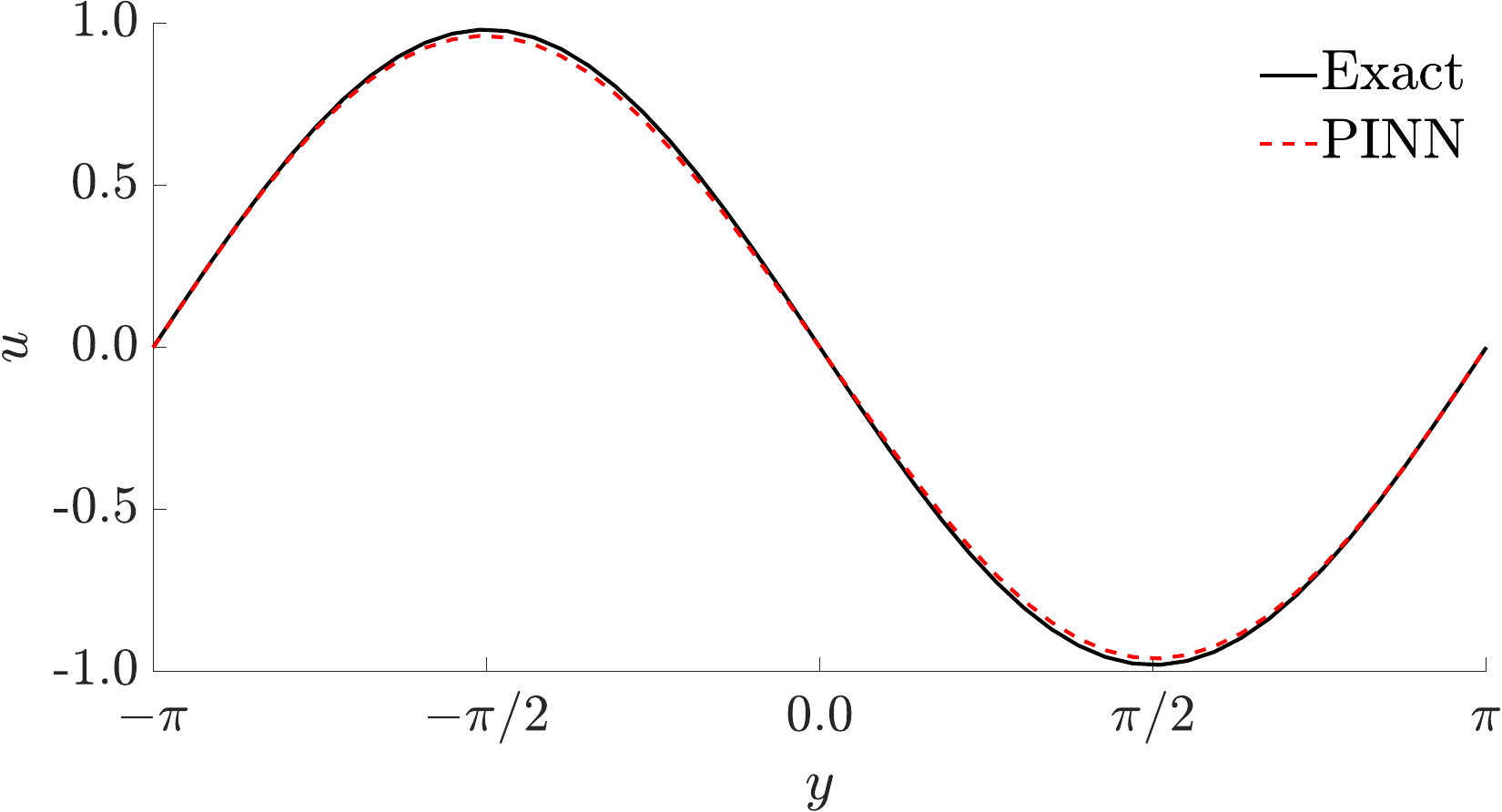}
        \caption{$t=1$}
    \end{subfigure}
    ~
    \begin{subfigure}[b]{0.45\textwidth}
        \includegraphics[width=\textwidth]{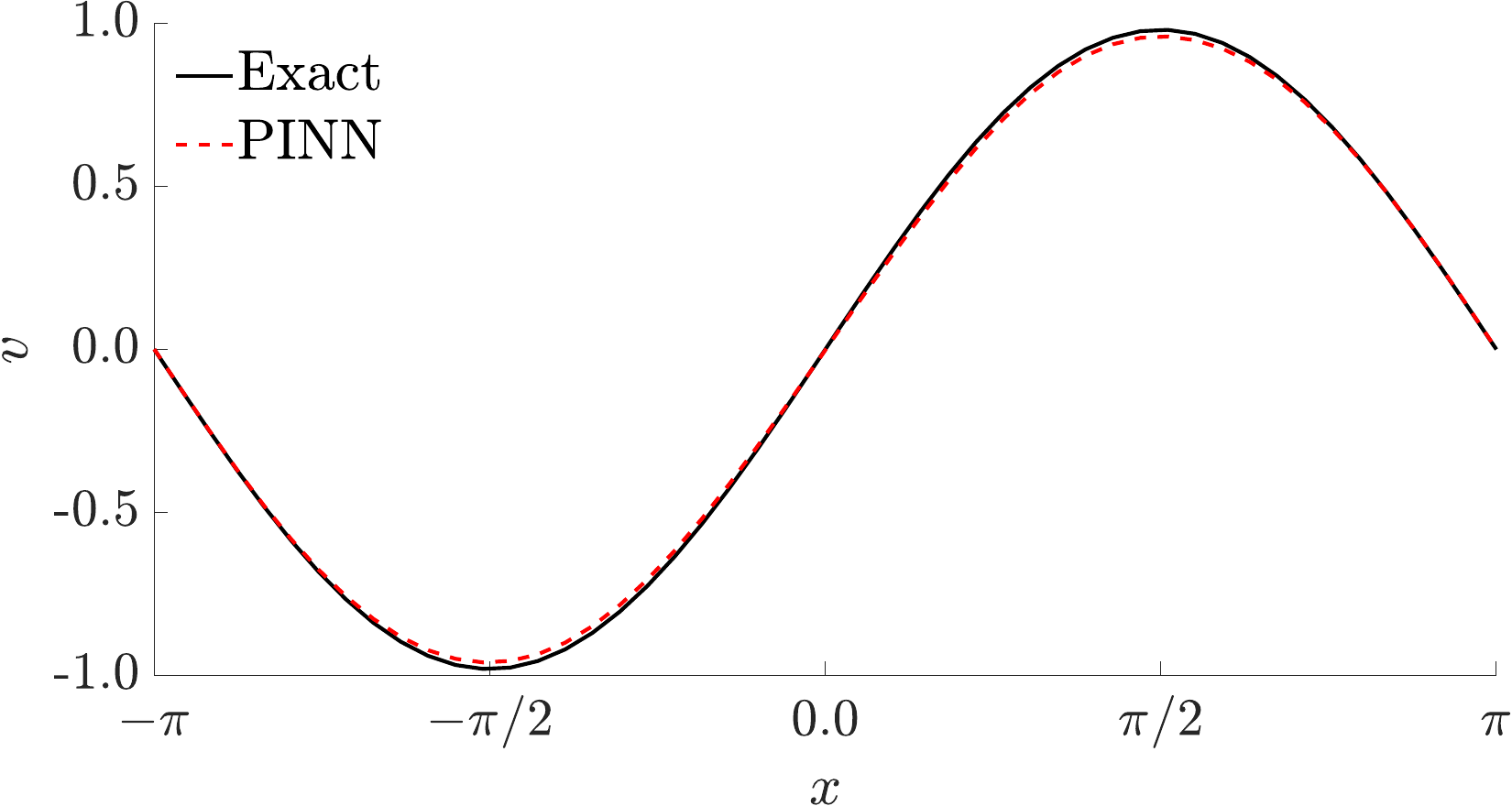}
        \caption{$t=1$}
    \end{subfigure}
    ~
    \begin{subfigure}[b]{0.45\textwidth}
        \includegraphics[width=\textwidth]{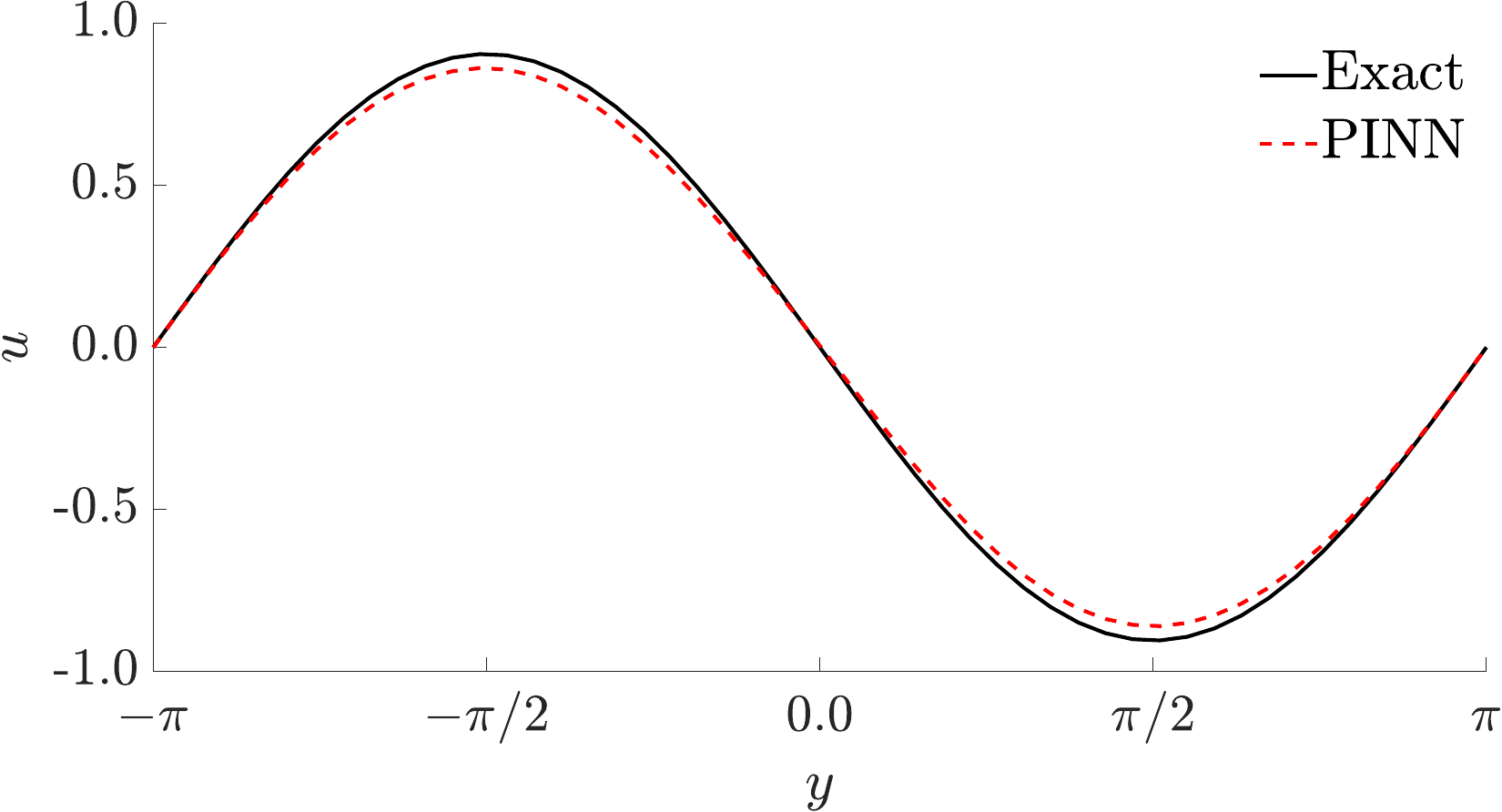}
        \caption{$t=5$}
    \end{subfigure}
    ~
    \begin{subfigure}[b]{0.45\textwidth}
        \includegraphics[width=\textwidth]{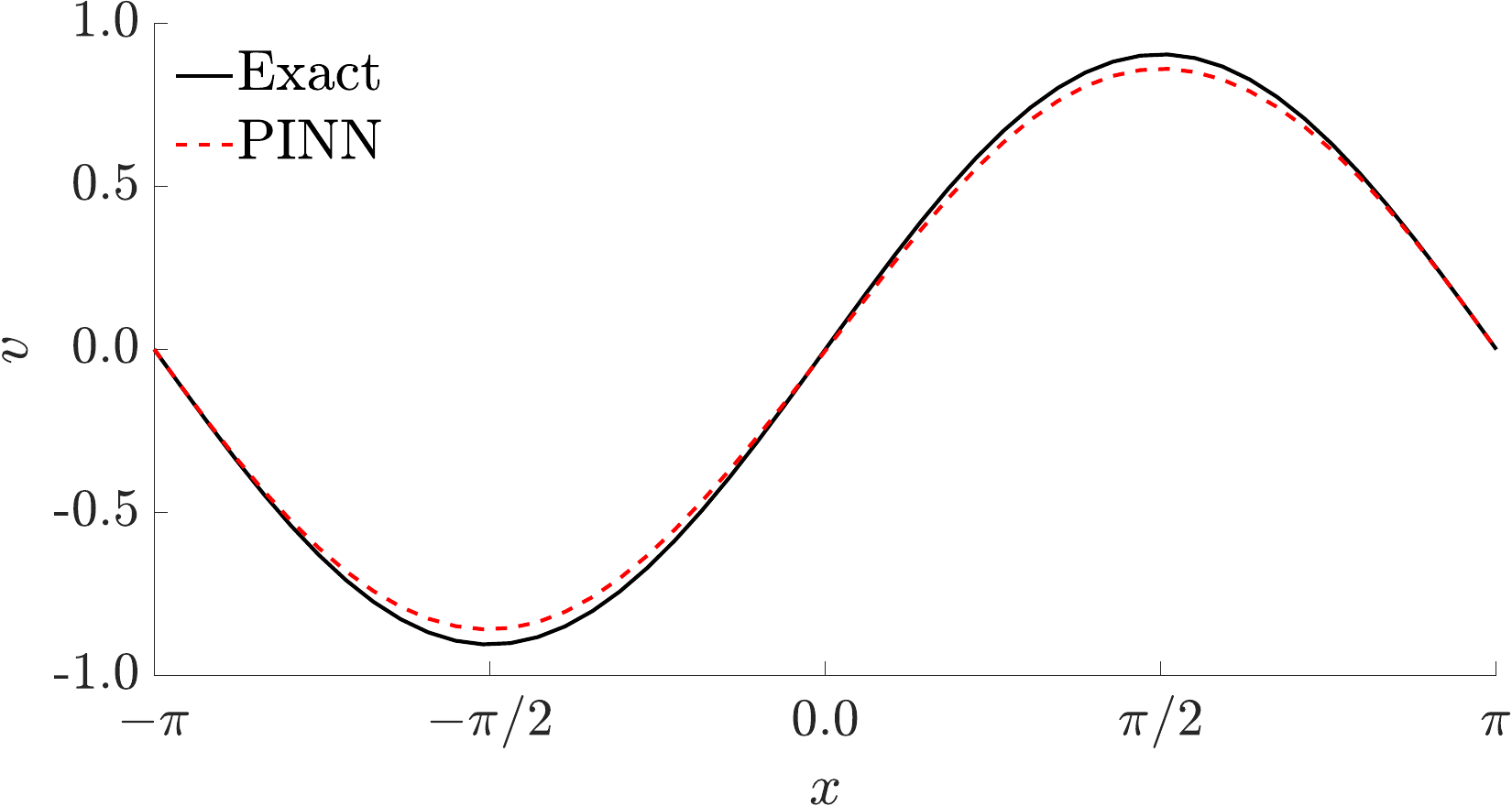}
        \caption{$t=5$}
    \end{subfigure}
    ~
        \begin{subfigure}[b]{0.45\textwidth}
        \includegraphics[width=\textwidth]{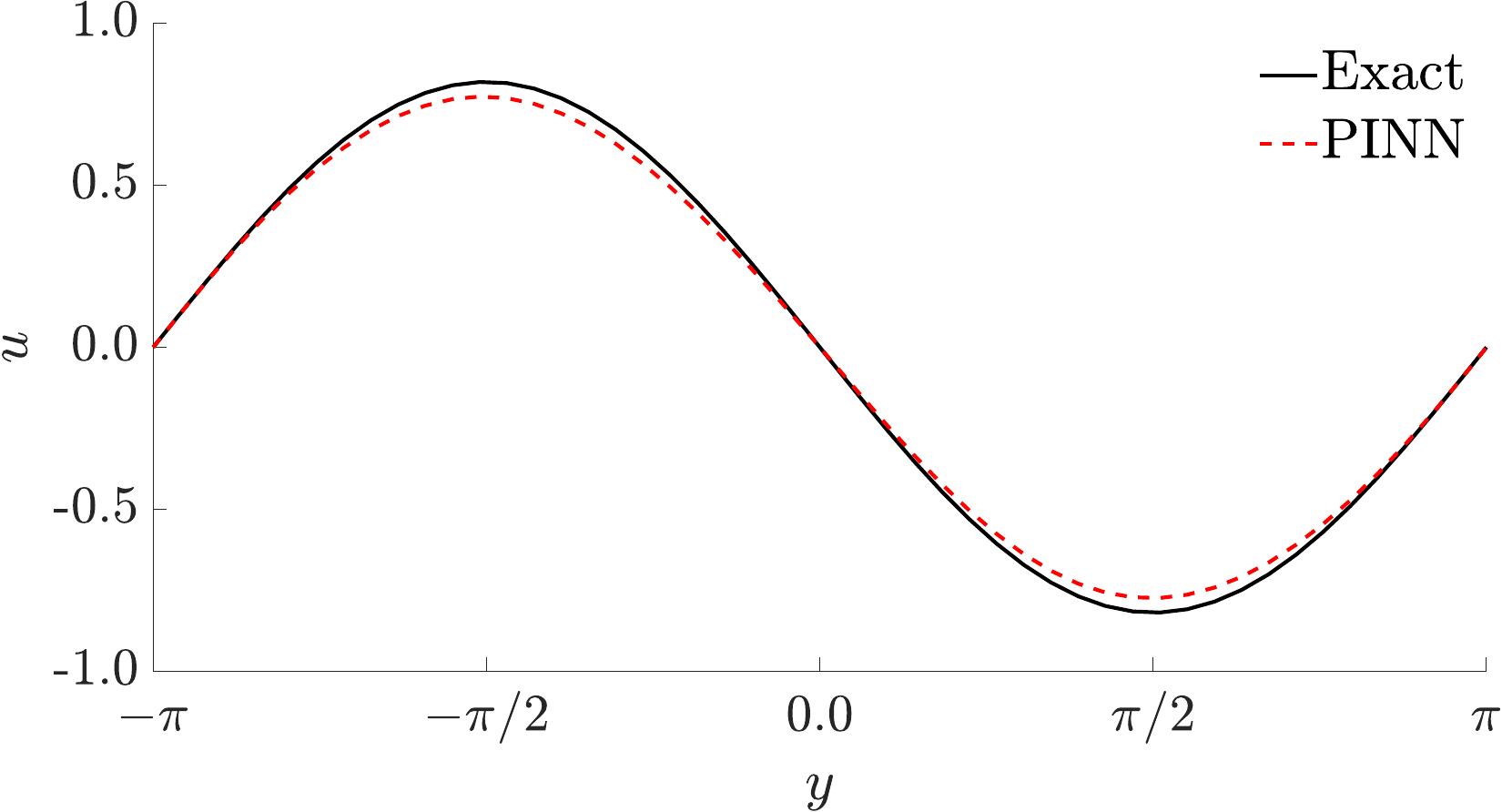}
        \caption{$t=10$}
    \end{subfigure}
    ~
        \begin{subfigure}[b]{0.45\textwidth}
        \includegraphics[width=\textwidth]{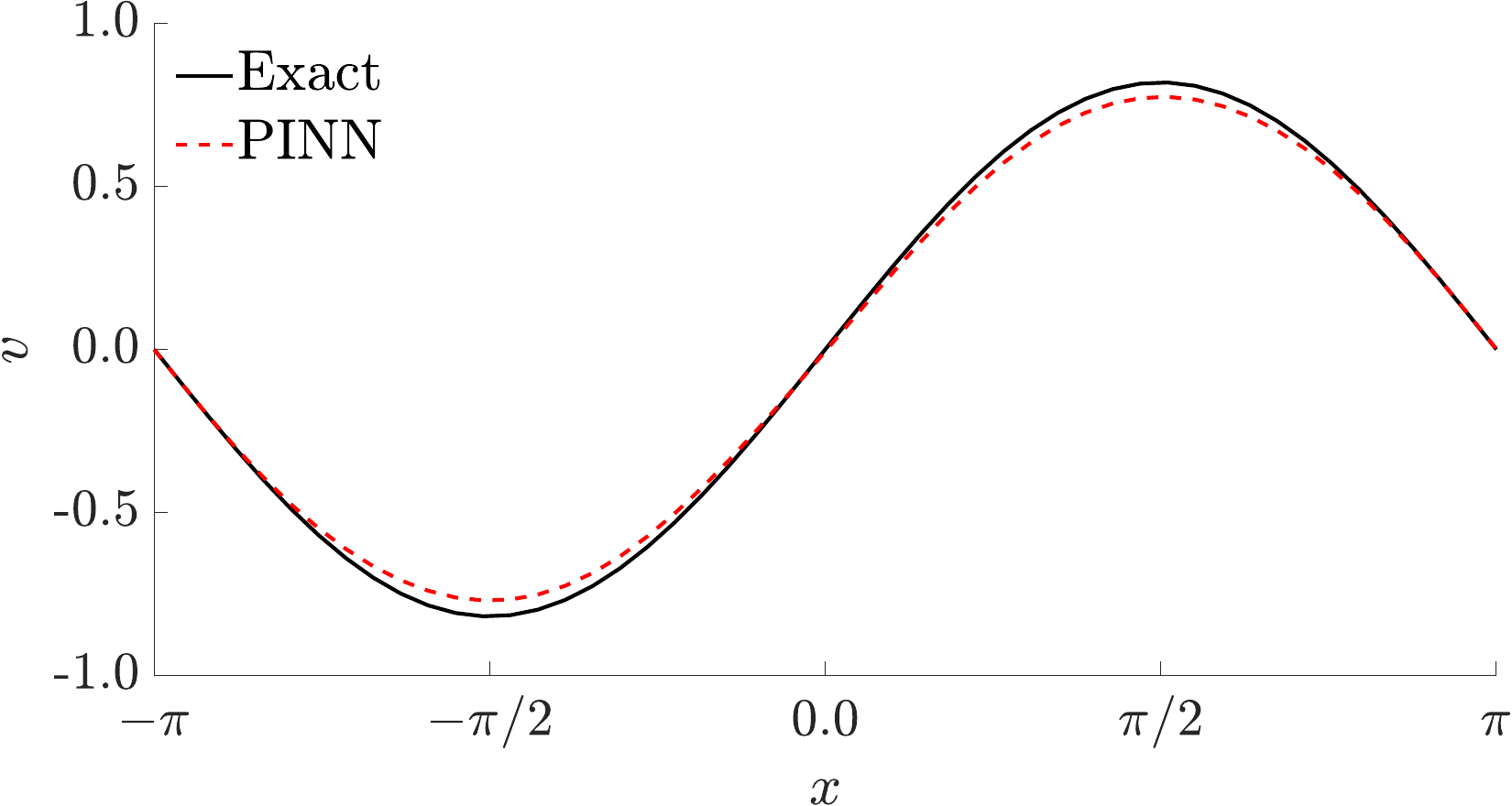}
        \caption{$t=10$}
    \end{subfigure}
    ~
    \caption{Velocity profiles at different time instances for Taylor-Green flow. The left column shows the $u$ velocity profile at $x=0.0$, and the right column shows the $v$ velocity profile at $y=0.0$.}
    \label{fig:tgv_vel}
\end{figure}

For the PINN solution, we again used two networks with 4 hidden layers and 128 neurons in each layer. We sampled 20000 points in the spatiotemporal domain and 1000 points in each boundary and initial time. We set $RT = 100$ and $\nu = 0.01$ for the unsteady simulations in $t \in [0, 10]$. We have used Equation (\ref{eq:tgv_exact}) for the boundary and initial conditions of the velocities. We have initialized the field with $\rho=1$. The initial learning rate is set to $1\times 10^{-3}$ and exponentially decays with a decay rate of 0.99. Both networks are initialized with the Glorot scheme and trained for 50000 epochs. We have presented the evaluation of normalized $L_2$ errors of $u$ velocity in Figure \ref{fig:Error_u_tgv} through the time domain. The error is smaller at times closer to the initial condition and increases as the information propagates in time \citep{wang_respecting_2024}. We have also provided centerline velocity profiles compared with the exact solutions in Figure \ref{fig:tgv_vel}. This demonstrates the accuracy of PINN using Galerkin-Boltzmann equations in time-dependent problems.

\subsection{Flow Over a Square Cylinder}
\begin{figure}
    \centering
    \includegraphics[width=0.7\linewidth]{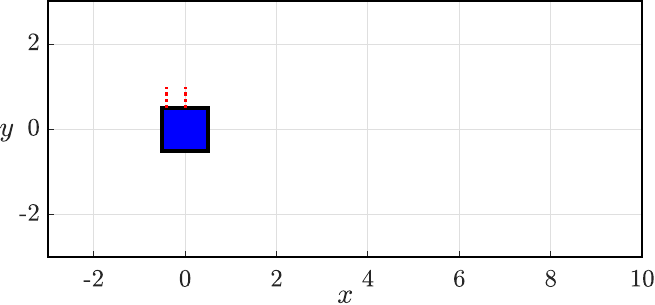}
    \caption{Computational domain for the flow over a square test case. Red dashed lines show the locations of the analyzed boundary layer velocity profiles.}
    \label{fig:square_domain}
\end{figure}

In our next test case, we study the flow over a square as a forward and inverse problem. The computational domain is bounded by $-3.0 \leq x \leq 10.0$ and $-3.0 \leq y \leq 3.0$. The square center sits on (0,0) with a side length of 1. The uniform flow is from the inlet at $x=-3.0$ with a magnitude of 1.0. The Reynolds number is $Re=100$, the speed of sound is $c=10$, resulting in a Mach number of $Ma=0.1$, and the relaxation time parameter $\tau = 1\times10^{-4}$. We analyze the solution by comparing the accuracy of the boundary layer velocity profiles at two different locations at $x = -0.4$ and $x=0.0$ on the top surface of the square cylinder. The computational domain and the locations of the analyzed boundary layer velocity profiles can be seen in Figure \ref{fig:square_domain}.

\subsubsection{Forward Solution}

We have solved the uniform flow over a square cylinder test case using a limited number of data from a high-fidelity solver. We have used libParanumal \citep{ChalmersKarakusAustinSwirydowiczWarburton2020} in which the Galerkin-Boltzmann formulation is implemented using the discontinuous Galerkin method for providing the labeled data. The high-fidelity simulation is conducted with a mesh with $K=6360$ triangular elements and polynomial order $P=4$. A perfectly matched layer is utilized to suppress the unphysical incoming waves from the boundary.

For the PINN solution, we used networks with 4 hidden layers with 128 neurons. We sampled 5000 collocation points, 500 points on each outer boundary, and 100 points on all sides of the square. We also fuse 150 random labeled data from libParanumal. The training is performed for 25000 epochs in 35.44 minutes. 

\begin{figure}
    \centering
    \begin{subfigure}[b]{0.45\textwidth}
        \includegraphics[width=\textwidth]{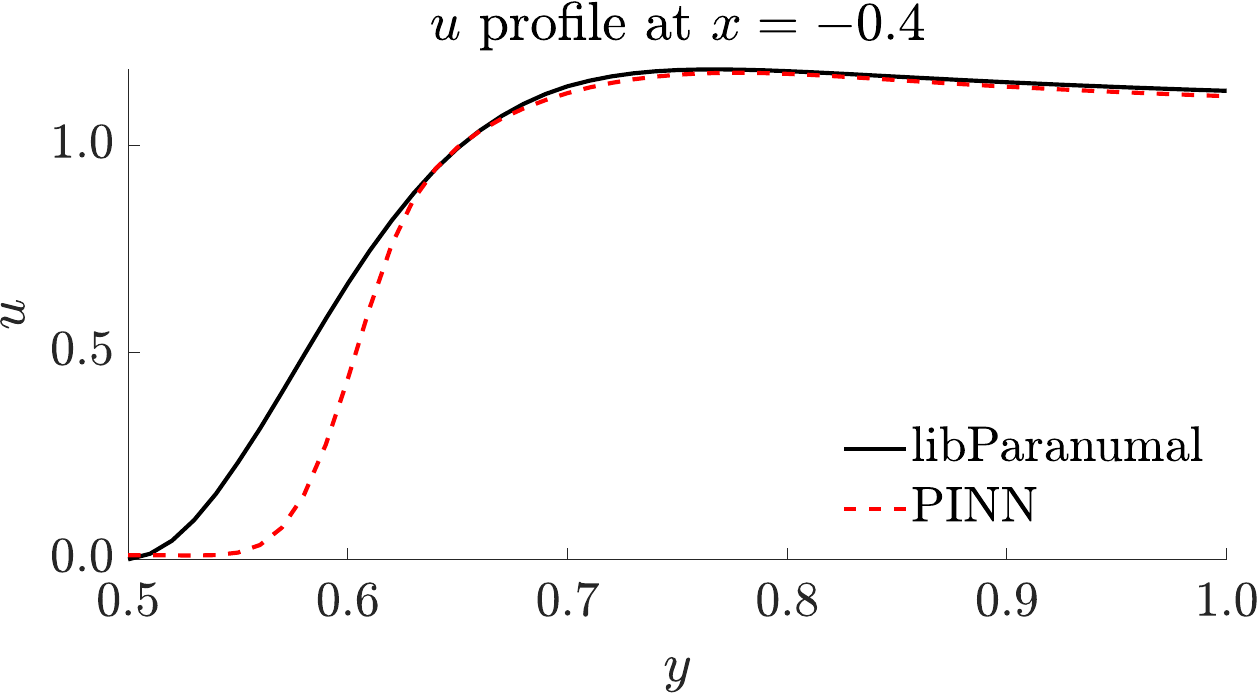}
    \end{subfigure}
    ~
    \begin{subfigure}[b]{0.45\textwidth}
        \includegraphics[width=\textwidth]{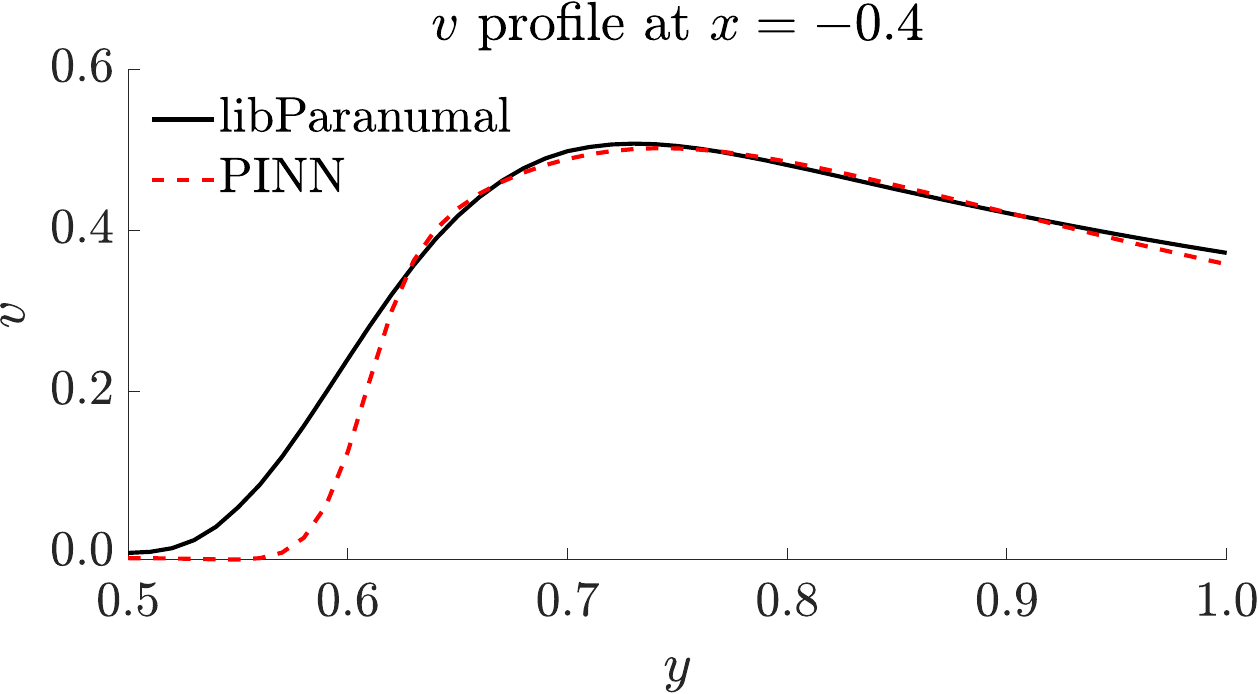}
    \end{subfigure}
    ~
    \begin{subfigure}[b]{0.45\textwidth}
        \includegraphics[width=\textwidth]{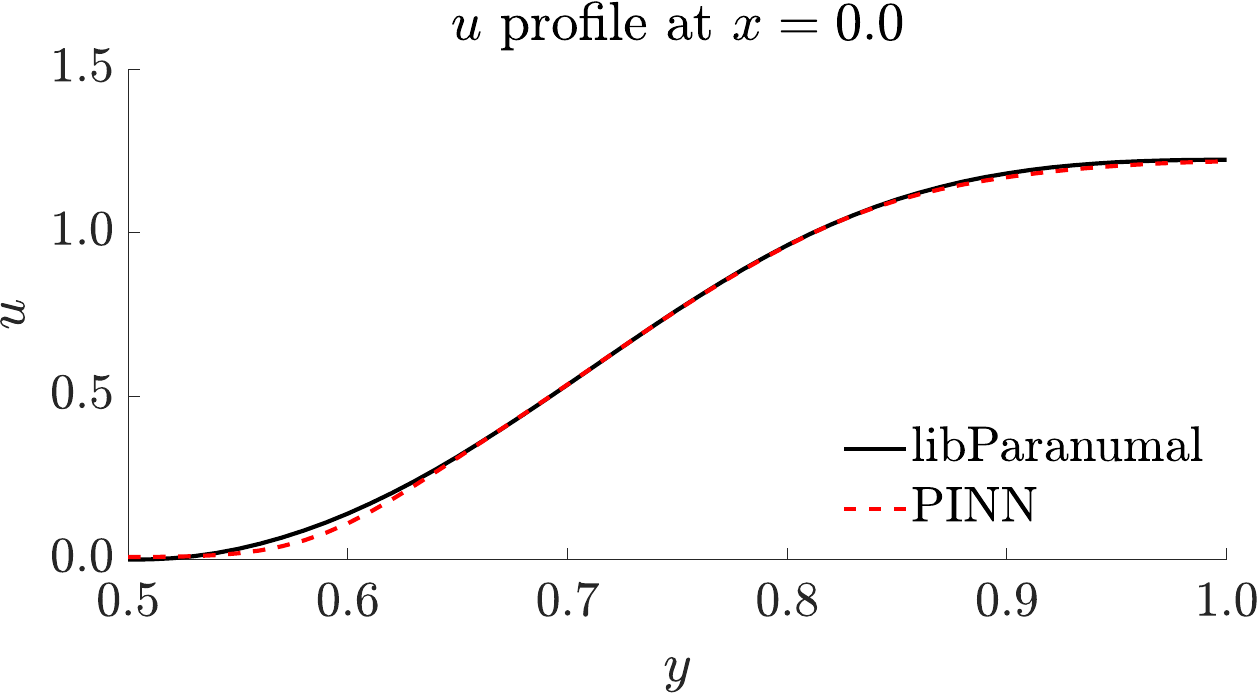}
    \end{subfigure}
    ~
    \begin{subfigure}[b]{0.45\textwidth}
        \includegraphics[width=\textwidth]{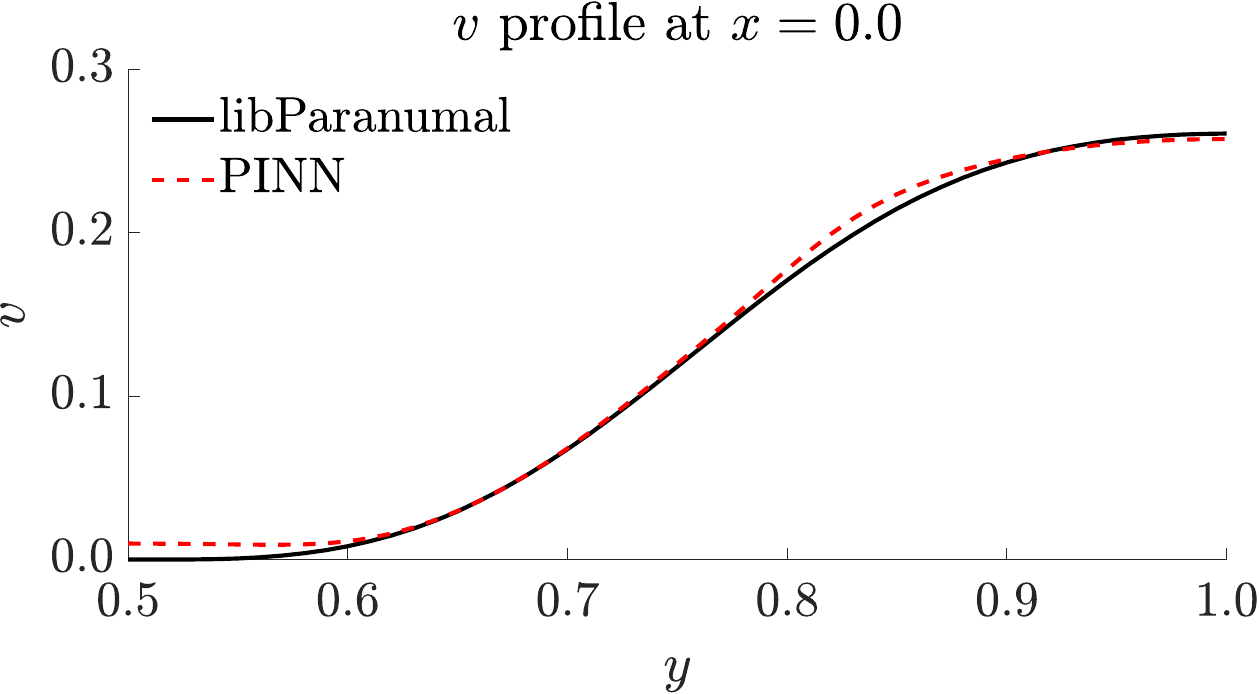}
    \end{subfigure}
    ~
    \caption{$u$ and $v$ velocity profiles at the boundary of the square. The velocity profiles start at the top boundary of the square at $y=0.5$. The top row shows the profile at $x=-0.4$, and the bottom row shows the profile at $x=0.0$.}
    \label{fig:square_BL}
\end{figure}

We compare and analyze the boundary layer profile at the top of the square using the high-fidelity data and PINN solution. In Figure \ref{fig:square_BL}, we present the $u$ and $v$ boundary velocity profiles between $0.5 \leq y \leq 1.0$ at $x=-0.4$ and $x=0.0$. The corner of the square is at $x=-0.5$, and therefore, the flow separates from the wall. The line at $x=-0.4$ is a high-gradient region, and PINN struggles to find an accurate solution, especially near the boundary. However, the solution can follow a trend similar to the high-fidelity solution obtained by libParanumal. At the boundary, the flow variables are not exactly satisfied since we implemented soft boundary conditions. At $x=0.0$, the middle point of the square, the gradient is lower than the corner of the square, and PINN can capture the velocity field accurately. 

\begin{figure}
    \centering
    \begin{subfigure}[b]{0.45\textwidth}
        \includegraphics[width=\textwidth]{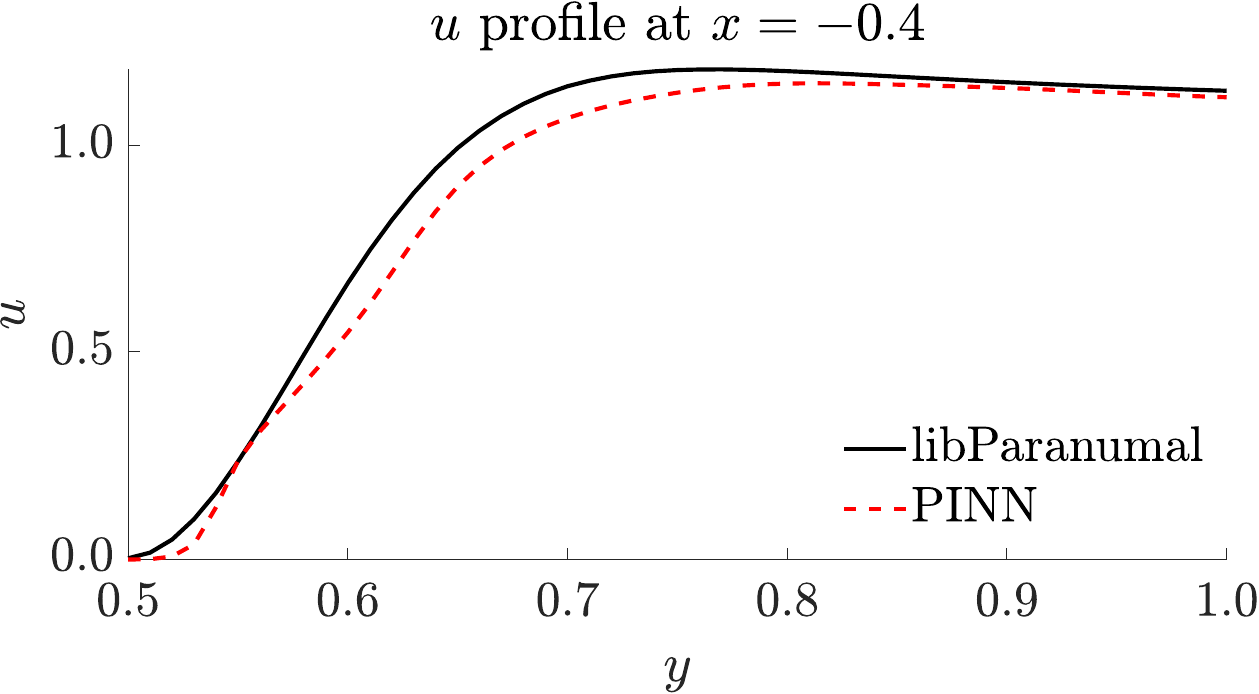}
    \end{subfigure}
    ~
    \begin{subfigure}[b]{0.45\textwidth}
        \includegraphics[width=\textwidth]{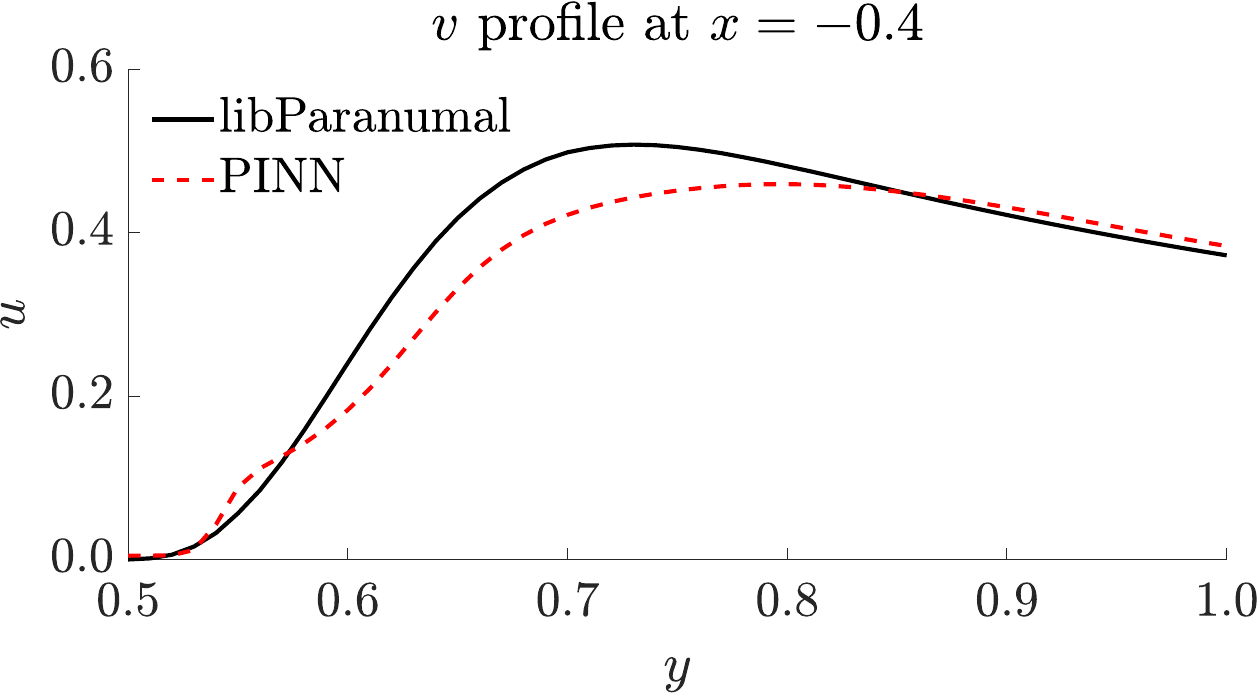}
    \end{subfigure}
    ~
    \caption{$u$ and $v$ velocity profiles at the boundary of the square with an increased number of collocation points near the boundary. The velocity profiles start at the top boundary of the square at $y=0.5$. The top row shows the profile at $x=-0.4$, and the bottom row shows the profile at $x=0.0$.}
    \label{fig:square_BL_dense}
\end{figure}

To enhance the performance of PINN to capture the boundary layer profile more precisely we have tried increasing the collocation point density around the square. We have created a zone in $-1.0 \leq x \leq 1.0$ and $-1.0 \leq y \leq 1.0$ and sampled 1000 points inside this zone, excluding the square boundary. We sampled 4000 points from the remaining domain, and the number of boundary samples is the same. The result is presented in Figure \ref{fig:square_BL_dense}. It can be seen that we can achieve an accurate boundary layer profile compared to the uniform sampling through the whole domain, especially in the near region of the boundary.

\subsubsection{Inverse Solution}

For the same test case, we have also tested the performance of this formulation in an inverse problem. We focus on predicting the relaxation time parameter with the high-fidelity data from libParanumal. Since this parameter is a very low number, we scaled the trainable parameter by an order of $10^4$. This way, the neural network predicts 10000 times $\tau$. The order is rescaled to its original order of magnitude in the PDE. The neural networks are trained for 25000 epochs with an initial learning rate of $5\times10^{-3}$ with an exponential decay rate of 0.99. 5000 labeled data points are used with an additional 5000 collocation points to satisfy the PDE. We assign 10 times higher weight to the residual loss to increase the learning towards the PDE along with the data loss. The PINN predicts the parameter as $1.05 \times 10^{-4}$, which is very close to the true value of parameter $1.0 \times 10^{-4}$.

%% file: conclusion.tex
In this work, we presented the Galerkin-Boltzmann formulation in a PINN framework and used it to simulate flows in weakly compressible regimes. The Galerkin-Boltzmann formulation with the BGK collision model is discretized with second-order Hermite polynomials in velocity space, leading to a first-order system of equations. Reducing the order of the derivatives compared to the Navier-Stokes equations and the number of equations compared with the widely used D2Q9 lattice Boltzmann formulation makes this system well-suited for PINN architectures. We created two different neural networks to separate the training of equilibrium and non-equilibrium states to overcome the scale discrepancy between the advection and the collision terms. 
We solved the Kovasznay flow using the Galerkin-Boltzmann formulation and the incompressible Navier-Stokes equations in PINN. The Galerkin-Boltzmann formulation results in low $L_2$ errors in velocity predictions in a comparable training time. We also presented the accuracy of the solver in unsteady problems using the Taylor-Green vortex. We analyzed boundary layer profiles of flow over a square test case by providing limited data from a high-fidelity solver. PINN struggles to predict accurate boundary layer velocity profiles in the regions where the flow is separated. However, increasing the collocation point density can overcome this challenge and provide more accurate solutions. These numerical tests show promising results in predicting flows in weakly compressible regimes.